# The Electron Capture Hypothesis – A Challenge to Neuroscientists

By John Robert Burger
Professor Emeritus
Department of Electrical and Computer Engineering
25686 Dahlin Road
Veneta, OR 97487
(jrburger1@gmail.com)

*Abstract* – Lower speed impinging ions (with hydration shells) cannot transverse ion channels once internal charge goes positive. Yet neural pulse waveforms fail to show the expected risetime distortion beginning at zero voltage. Observed waveforms cannot be explained unless electron capture is considered.

## Introduction

Within an ion channel lower speed positive ions are stopped a reversed by the repelling electric field as soon as internal voltage goes positive. A typical potential barrier approaches +40 mV; assuming a 4 nm long channel this amounts to +10 megavolts per meter (MeV/m) which is very high indeed, higher than the field for lightening in the sky. Reversed ions drift back to where they came from and obviously do not transfer charge. Consequently there ought to be a reduction in the risetime of a neural pulse as internal voltage goes positive. Such distortions are not observed. To resolve this dilemma, it has been proposed that instead of ions being transmitted, electrons are captured (Burger 2009).

The strongest argument against electron capture is that ion penetrations into neurons during action potentials apparently explain energy usage within the brain (Attwell 2001). A fashionable hypothesis is that in order to restore ionic balance within spiking neurons ATP pumping consumes energy. Experiments have tried to prove that action potentials are driven by ion penetration into neurons (Keynes 1951) but not everyone agrees with the conclusions of such experiments. In the interests of scientific skepticism this paper ignores ion channel dogma and considers a conceivable alternative.

## The Waveform Dilemma

A typical waveform for a neural pulse as observed on an oscilloscope is illustrated in Fig. 1. It rises from about -55 mV up to about +40 mV on a fairly straight line. The slope of the rise is about constant, which is characteristic of a constant current source charging fixed membrane capacitance. The constant current source is assumed to be generated by the thermal activity of ions impinging on transmembrane proteins known as ion channels.

By way of review, neural membranes are surrounded inside and out by ionic solutions. Ions in solution have hydration shells (of about five polar water molecules) and a distribution of velocities. In this case average speed is about 238 m/s.[1]. These velocities are based on calculations assuming the mass of a sodium ion plus five water molecules. Fig. 2 is a typical probability distribution of kinetic energies at 310 K (http://en.wikipedia.org/wiki/Maxwell%E2%80%93Boltzmann_distribution). Note that energy in meV (millielectron-volts) is equal to voltage in mV (millivolts).

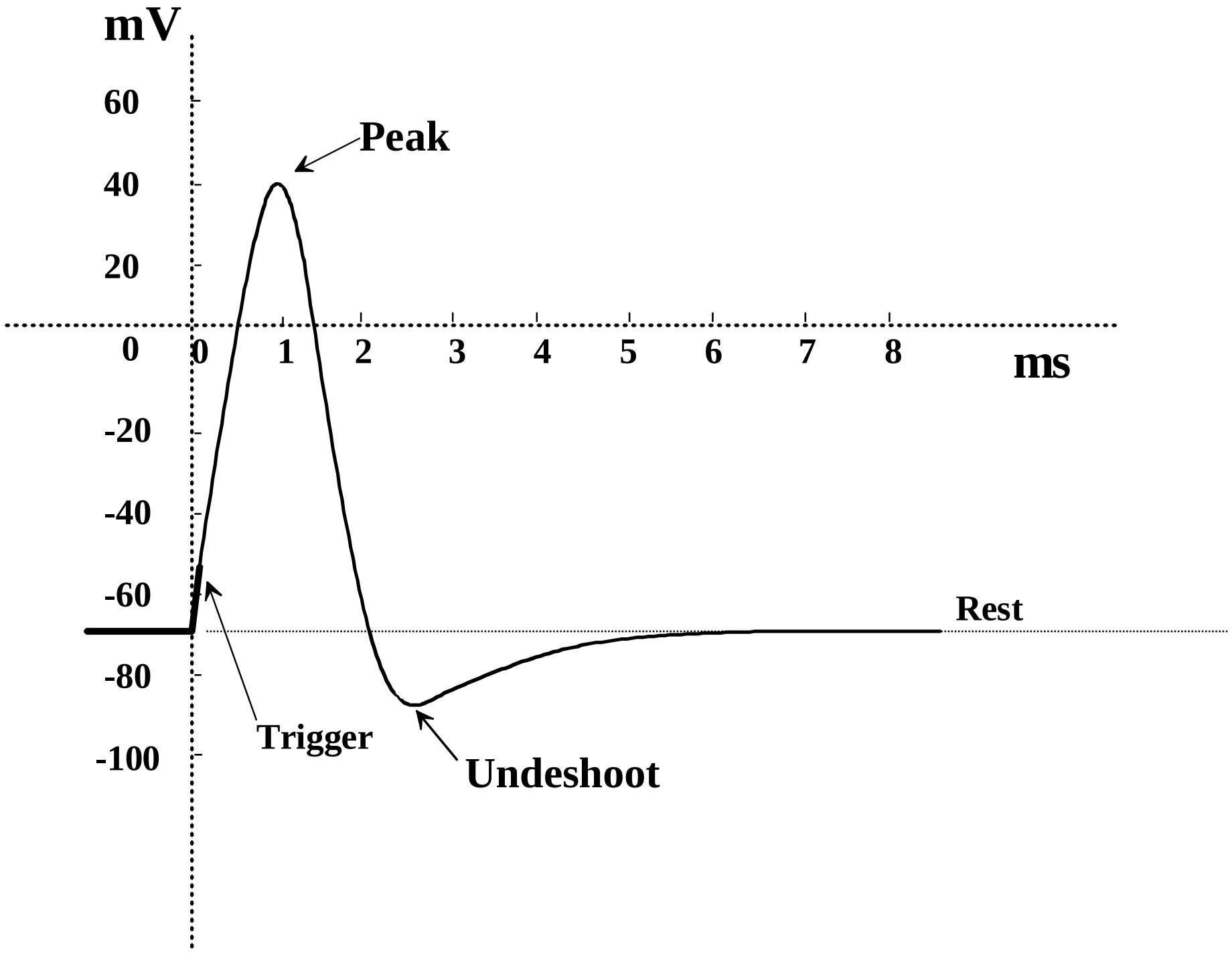

**Fig. 1 Typical Neural Pulse**

When an ion diffuses into an open channel, it undergoes a pull from the negative field and thus drifts through the channel and enters into the neuron. Neglecting the possibility of quasi-free electrons in the channel, all such ions diffusing into the channel are pulled completely through. When interior potential goes positive relative to the outside, clearly not all entering particles go through. Those with lower speed ($v$) are repelled. It can be shown that the penetration of a charged particle into a repelling electric field is $E_k/E$ where $E_k$ is the kinetic energy $(1/2\ Mv^2)$ expressed in electron volts ($eV$) and where $E$ is the opposing electric field in volts per meter ($V/m$) (Hemmenway 1962). Fig. 3 plots a typical electric field across a hypothetical 4 nm ion channel in a neural membrane.

Note that mass $M$ affects speed but not energy. A particle gets through a channel of length $L$ if

$$\frac{E_k}{E} = \frac{E_k}{V/L} > L$$
$$\Rightarrow E_k > V$$

---

[1] $\langle v\rangle = \sqrt{\frac{8kT}{\pi M}} \approx 238.5\ m/s$; *Mass M for Na + 5 $H_2O$ is calculated to be 18.83 x $10^{-26}$ kg; T = 310 K; k = 1.38 x $10^{-23}$ J/K*

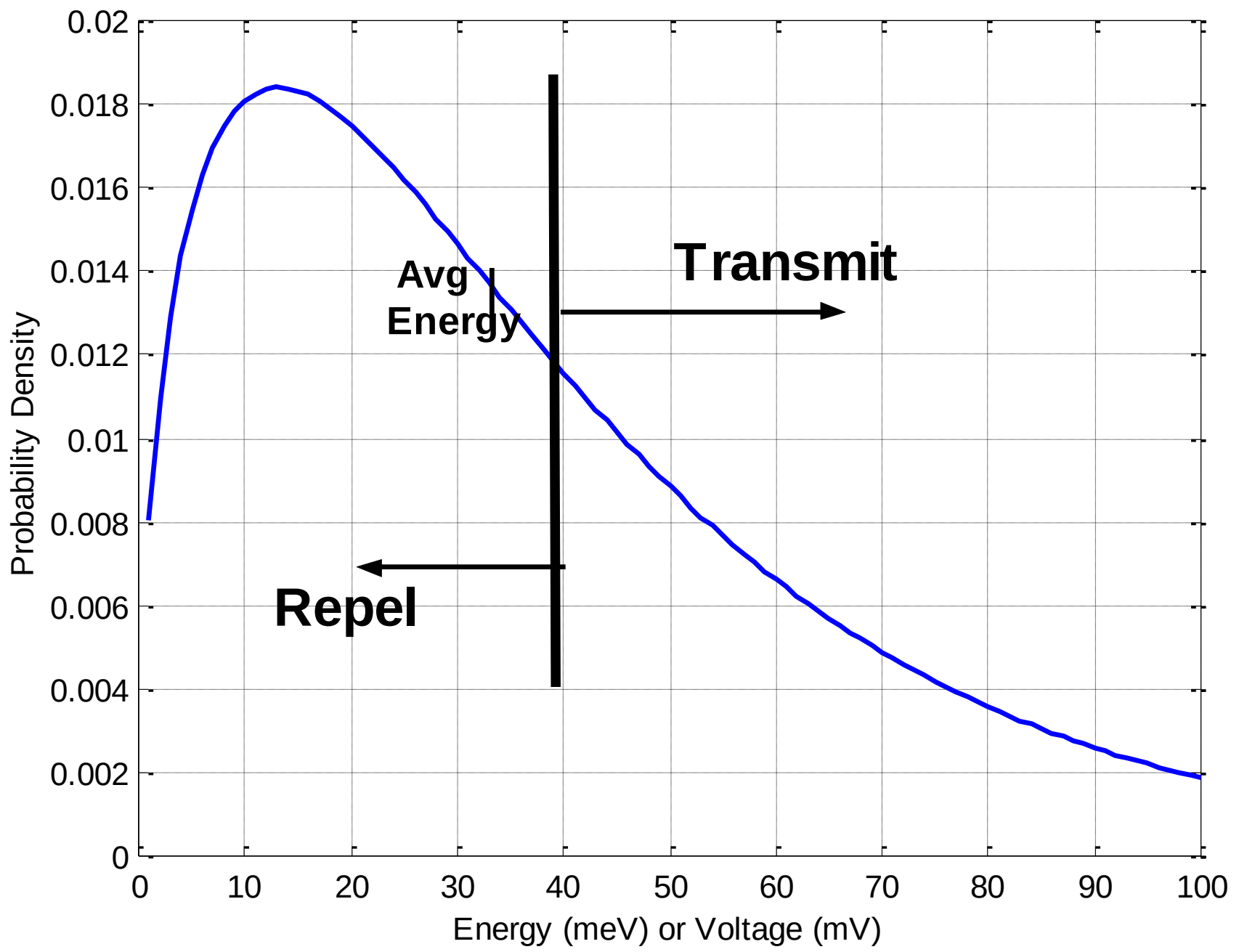

**Fig. 2 Thermal energy distribution (Repel if $E_k < V$; Transmit if $E_k > V$)**

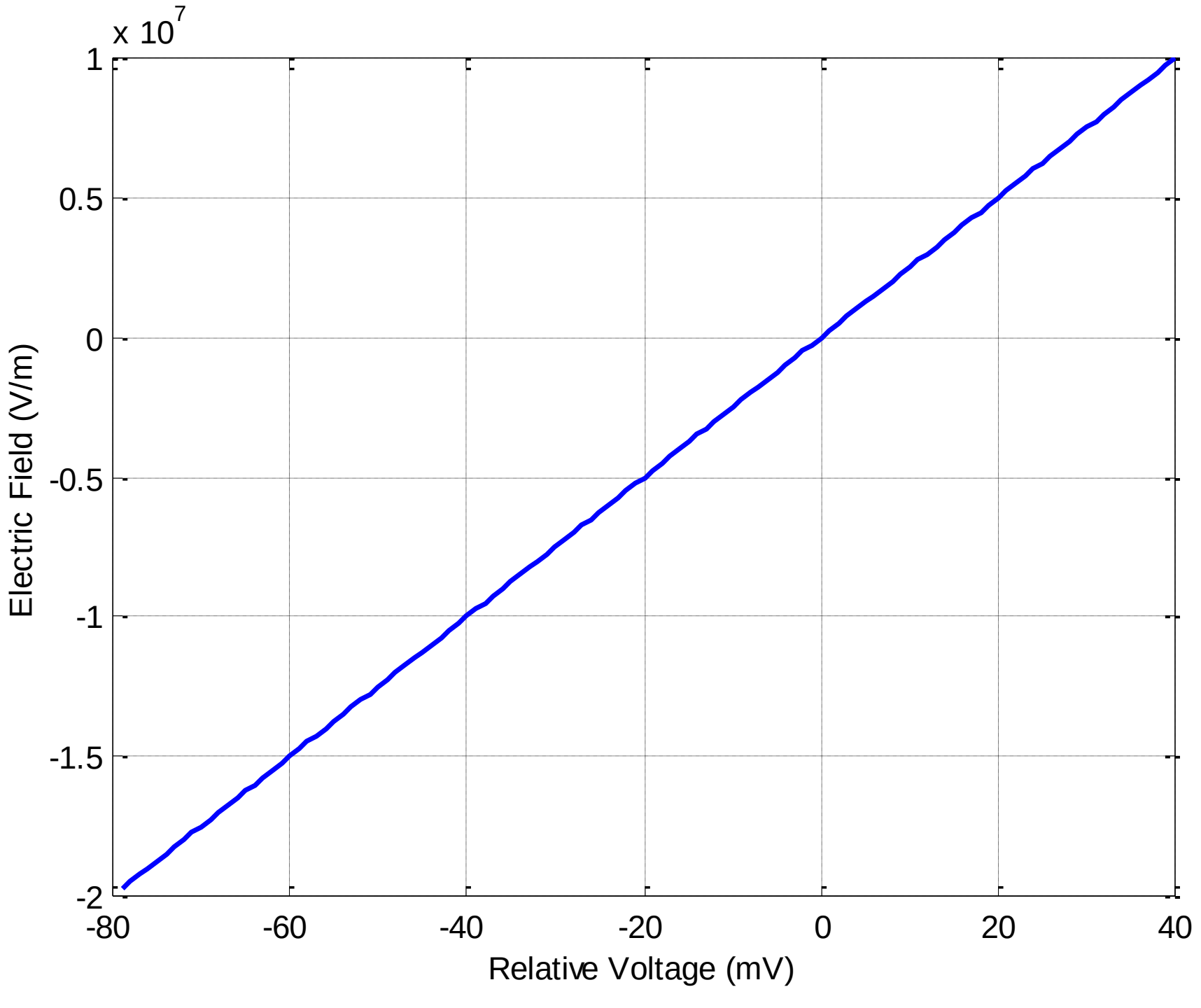

**Fig. 3 Electric Field in Ion Channel as a Function of Membrane Voltage**

If the energy of a particle $E_k$ is numerically below the accumulated interior voltage $V$, transmission is impossible. Note that membrane wall thickness L is not directly involved in whether or not a particle gets through, because field $E$ in *V/m* is inversely proportional to $L$. All that matters is particle speed (kinetic energy $E_k$) and interior voltage $V$.

The average kinetic energy due to thermal activity is equal to

$$\langle E_k \rangle = \frac{1}{2} M \langle v \rangle^2 = \frac{4}{\pi} kT \approx (1.273)(1.38x10^{-23})(310) = 5.447x10^{-21} J = 34.0\ meV$$

As neural voltage builds up toward its peak of about 40 mV, cumulative probability becomes quite large that ions will be repelled as suggested in Fig. 2. Where do they go? Mostly they reverse directions and leave.

If particle B with speed $u_B$ tries to enter before particle A with speed $u_A$ leaves, a collision is expected since ion channels are usually envisioned as somewhat confined. Assuming an ideal elastic collision, momentum is conserved. For example, if $u_A = 0$ while $u_B$ has higher than average energy, particle A will be knocked through the channel while particle B will be stopped (http://en.wikipedia.org/wiki/Momentum). It is highly unlikely that a stream of successive ions will have energies higher than <E>. Statistically most entering ions have lower energies than <E> according to the energy distribution, and are repelled, reducing ionic current.

Significantly, negative charges attached to the walls of a channel cannot be ignored. For instance, negative charges would repel chlorine anions. But they also attract sodium ions. So sodium is further slowed by local charges.

The number of potential ion channels (P) is expected to remain limited by the number of transmembrane proteins per $cm^2$, roughly $10^{10}$ sodium channels/$cm^2$, according to photographs of the membrane (http://www.cytochemistry.net/Cell-biology/membrane_intro.htm). It has been estimated that inrush ion current is $J_{ION} \approx 1.8\ x\ 10^{15}$ ions/s/$cm^2$ corresponding to an average inrush current of 0.3 mA/$cm^2$ through a patch of active membrane. This translates to $J_{ION}/P = 1.8\ x\ 10^5$ ions/s through each channel, or one charged ion about every 5.5 us during normal charging. However, if channels become blocked, charging rate cannot be this high.

Without electron capture the neural pulse might appear at in Fig. 4. Thus enters the hypothesis that electron capture is occurring. Electrons in ionic solutions are quasi-free for only nanoseconds, but because of their small mass, 1/1836 of a proton, they achieve high velocities; they travel far (Conway 1981). They will penetrate into as opposing electric field as readily as any other charged particle. Electrons are small enough to tunnel quantum mechanically through barriers where classically a particle cannot go. Plus they are easily captured. Their small size gives electrons an ability to go through physical barriers that block larger particles. The situation is envisioned in Fig. 5.

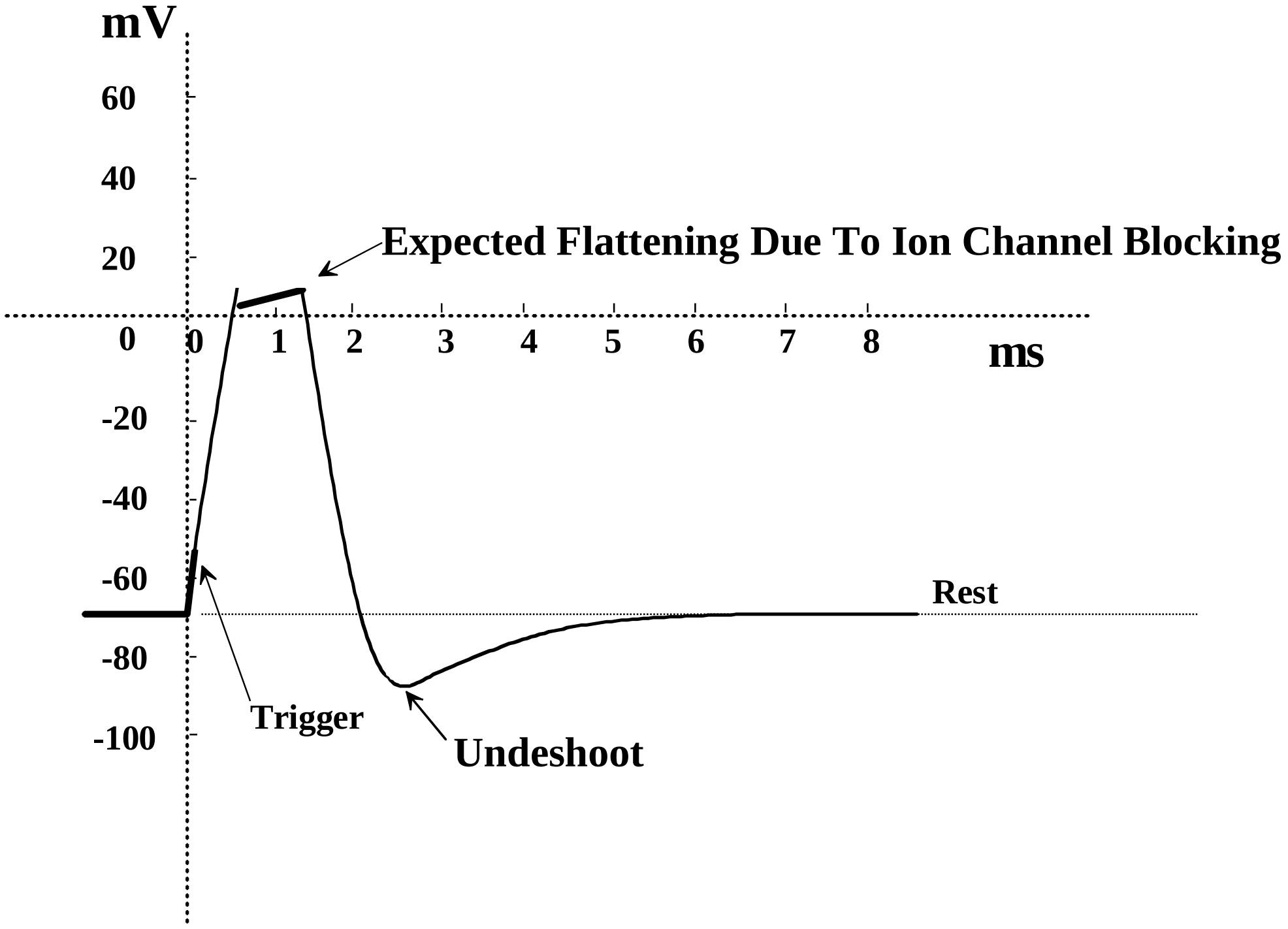

**Fig. 4 Expected Waveform Distortion**

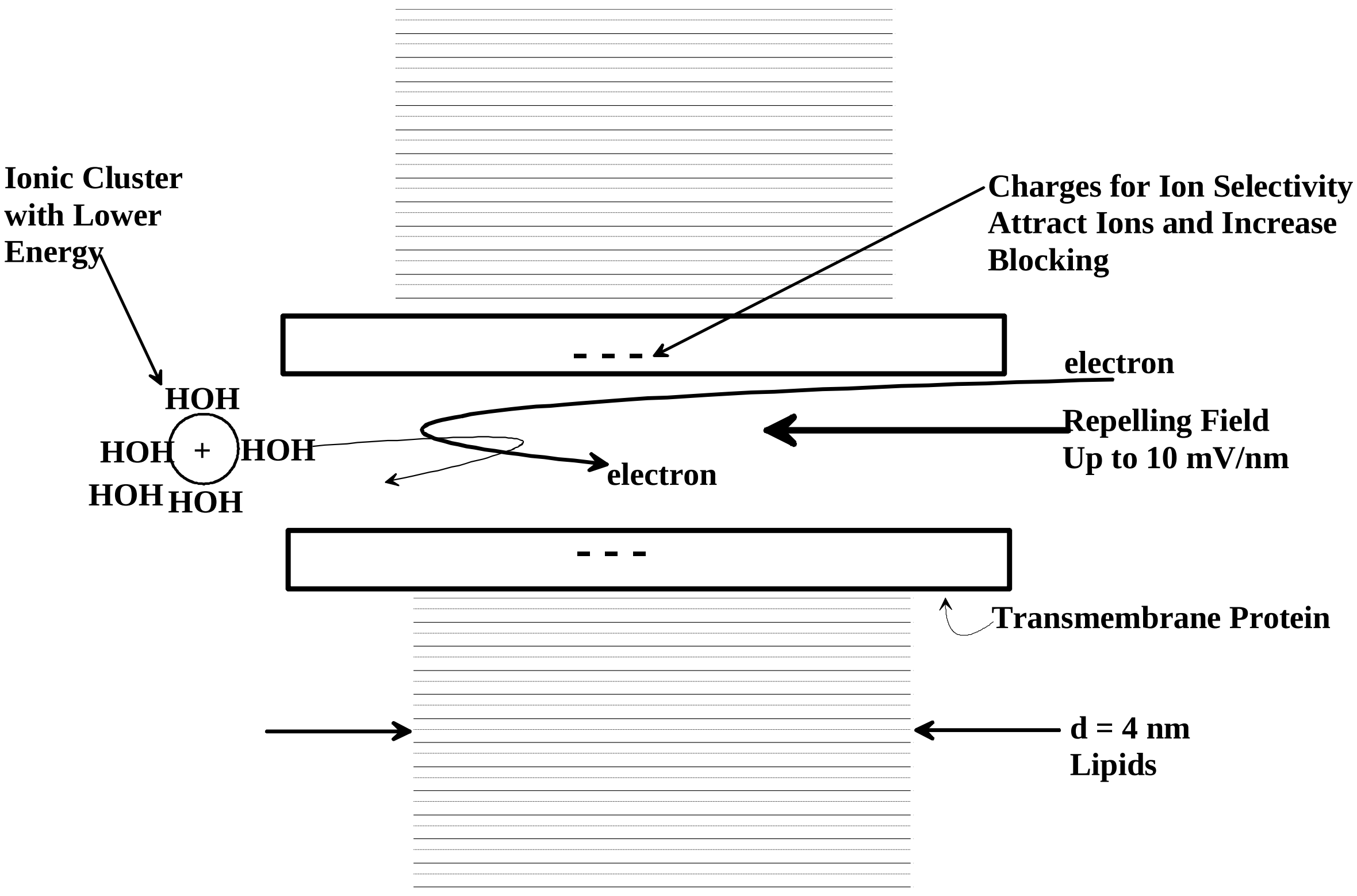

**Fig. 5 Schematic of a Blocked Ion Channel**

## Energy Dissipation in Neurons

Electron capture as a means of charge transfer would profoundly affect neural energy usage during signaling. This is because electrons may be reversible; they may be captured in a controlled way via relatively low currents, without dissipating a lot of $i^2R$ energy (Burger 2009). Controlled charging is actually suggested by the relatively slow rise and fall times of a neural pulse. It is much like the slow discharging and recharging of a rechargable battery. Entering ions, on the other hand, have to be pumped out using ATP pumps, and these pumps dissipate a lot of energy.

Fortunately, it would not be difficult to correct current theories. For instance, fMRI is supposed to image extra oxygenated hemoglobin pumped in for oxygen-consuming ATP pumps. Ions during charging enter at a rate of about $J_{ION} \approx 1.8 \times 10^{15}$ *ions/s/cm*. Each charging takes about 1 ms; there may be 100 charges per second. So for each second the average number of impurity ions that enter is $J_{ON}$ $(1 \times 10^{-3})$ (100) = $1.8 \times 10^{14}$ ions/s. Ion pumping occurs at purposeful transmembrane protein sites which are limited in number, perhaps $10^{10}/cm^2$. This makes $1.8 \times 10^{14}/10^{10} \approx 1.8 \times 10^4$ ions/s on average for each pump, which is incredibly large.

Assuming fairly continuous neural activity as in humans, ATP pumps indeed are very busy. Consequently vast energies are dissipated. The corresponding overcompensation by hemoglobin is what fMRI is thought to image (Ogawa 1990; Huettel 2004).

There are other possibilities. Neural membranes take continuous energy[2]. The very first major event during an action potential is an immediate reduction in the average voltage across the neural membrane. This results in an immediate and significant reduction in the rate of oxygen usage and a higher average of oxygenated hemoglobin. This, plus a lower level of parasitic ionic leakage and subsequent ATP pumping could account for fMRI images of neural activity.

By the way, properly applied patch clamps are commonly said to observe spikes due to ion channels opening and closing (http://en.wikipedia.org/wiki/Patch_clamp). Such wording leads one to believe that ions have been observed individually entering a neuron, one about every 5 us. Unfortunately, published oscillographs raise questions about the accuracy of this exaggerated claim. One problem is that such experiments do not distinguish between ions and electrons. Patch clamps certainly have not proved that electrons are irrelevant.

### Conclusions

Electron capture is vaguely similar to the induction of charge on a plastic comb that is brushed repeatedly through dry hair. For those who doubt the existence of electronic charge in water, the comb is readily seen to deflect a narrow stream of ordinary tap water. The point is, electrons are invisible yet powerful.

Modern science has nearly succeeded in convincing everyone that life depends on an inefficient leaky process in which charge is carried exclusively by positive ions entering a neuron. But such ions quickly poison the interior. They have to be pumped out before life expires, consuming vast amounts of energy. Electron capture, if it proves to be true, would greatly reduce this wasted energy. If true, neuroscience would again be compatible with Maupertuis' principle of least action: <u>Nature is thrifty in all its actions</u>.

---

[2] Assume a rest voltage of -70 mV and a membrane conductance of 0.3 $mS/cm^2$:
$P_{LOSS}=0.07^2\ 0.0003 = 1.47\ uW/cm^2$
Assume a spherical neural membrane of diameter D = 25 um:
$E_{MEM} = (1.47\text{x}10^6\ W/cm^2)\ (\pi D^2)\ (10^2\ cm/m)^2\ (1\ kcal/4.184\ kW)\ (1\ kW/10^3W)\ (3600\ s/h) = 2.48 \times 10^{-11}\ kcal/h$